# The inconsistency of h-index: a mathematical analysis

Ricardo Brito[a*], Alonso Rodríguez Navarro[a,b]

[a] *Departamento de Estructura de la Materia, Física Térmica y Electrónica and GISC, Universidad Complutense de Madrid, Plaza de las Ciencias 3, 28040, Madrid, Spain*
[b] *Departamento de Biotecnología-Biología Vegetal, Universidad Politécnica de Madrid, Avenida Puerta de Hierro 2, 28040, Madrid, Spain*

*\* Corresponding author e-mail address: brito@ucm.es*

*A R-N e-mail address: alonso.rodriguez@upm.es*

**Abstract**

Citation distributions are lognormal. We use 30 lognormally distributed synthetic series of numbers that simulate real series of citations to investigate the consistency of the *h* index. Using the lognormal cumulative distribution function, the equation that defines the *h* index can be formulated; this equation shows that *h* has a complex dependence on the number of papers (*N*). We also investigate the correlation between *h* and the number of papers exceeding various citation thresholds, from 5 to 500 citations. The best correlation is for the 100 threshold but numerous data points deviate from the general trend. The size-independent indicator *h/N* shows no correlation with the probability of publishing a paper exceeding any of the citation thresholds. In contrast with the *h* index, the total number of citations shows a high correlation with the number of papers exceeding the thresholds of 10 and 50 citations; the mean number of citations correlates with the probability of publishing a paper that exceeds any level of citations. Thus, in synthetic series, the number of citations and the mean number of citations are much better indicators of research performance than *h* and *h/N*. We discuss that in real citation distributions there are other difficulties.

## 1. Introduction

The *h* index, proposed by the physicist Jorge Hirsch (2005), has reached an astonishing popularity as an indicator of research quality. It has been applied to sets of publications at different level of aggregation, from a single author to a country. It is calculated by sorting the set of publications in descending order according to the number of citations, starting with 1 for the most cited publication. The *h* index is then the lowest rank of the publication that ranked at *h* has *h* or more citations. It is worth noting that the plot serving to determine the *h* index in this way corresponds to a rank/frequency plot (Newman, 2005). By varying the formulation bases, many variants of the *h* index have been proposed, but these variants do not offer clear advantages over the original *h* index (Bornmann, Mutz, Hug, & Daniel, 2011). Many studies have investigated mathematical formulations that calculate the *h* index using the number of citations and papers (e.g. Bertoli-Barsotti & Lando, 2017; Malesios, 2015). These publications address the relationship between the *h* index and other bibliometric parameters, but do not address the question of whether the *h* index can be used to assess research.

The popularity of the *h* index is revealed by the number of citations of the paper in which it was reported. From the date of its publication up to the time this study was performed (February 12, 2020), the Hirsch's paper (Hirsch, 2005) had received 4,405 citations in the Web of Science (WoS). Despite this impact, several studies have demonstrated that the *h* index is inconsistent and arbitrary (Schreiber, 2013; Waltman & van-Eck, 2012). Although "it appears therefore that a similar fate to that of the Journal Impact Factor awaits the *h*-index" (Bornmann, 2014, p. 750), the use of the *h* index is not decaying in citizen bibliometrics.

As described above, and like many other research performance indicators, the *h* index is based on citation counts. Present study assumes that, with the important limitations that we describe below, the number of citations reflects the scientific "value" of a publication. This matter has been extensively studied (e.g. Aksnes, Langfeldt, & Wouters, 2019; De-Bellis, 2009), and correlations of citation-based indicators with Nobel Prizes (Rodríguez-Navarro, 2011, 2016) and peer reviews (Rodríguez-Navarro &

Brito, 2020; Traag & Waltman, 2019) give support to this view. Weak correlations between citation indicators and peer reviews (Wilsdon et al., 2015) only mean that the tested indicators fail to match peer reviews but not that all citation-based indicators fail.

The *h* index was initially proposed "to quantify an individual's scientific research output" (Hirsch, 2005, in title), and this was the first error of the proposal because, formally, bibliometrics cannot be used for research assessments at low aggregation levels. There is general criticism of the use of citation counts for research assessments (MacRoberts & MacRoberts, 2018) and it is evident that there are too many cases in which the number of citations does not reflect the importance or the influence of a paper in its research field. This applies not only to the societal impact (Bornmann, 2013; Bornmann & Haunschild, 2017) but also to the scientific impact. Clear representatives of papers that are more important than their citation counts suggest are the "sleeping beauties" (van Raan, 2004), but they are not exceptions because "papers whose citation histories are characterized by long dormant periods followed by fast growths are not exceptional outliers, but simply the extreme cases in very heterogeneous but otherwise continuous distributions" (Ke, Ferrara, Radicchi, & Flammini, 2015, p. 7431). It has also been observed that many novel papers do not have an early recognition in terms of the number of citations (Wang, Veugelers, & Stephan, 2017).

This means that the number of citations does not measure the scientific relevance of a paper. It just correlates with its scientific importance. This is a completely different statistical concept that has been summarized as follows:

> Citation counts do not measure research impact under any reasonable definition of impact. Instead they are indicators of research impact or quality in the sense that in some disciplines, when appropriately processed, they tend to correlate positively and statistically significantly with human judgments of impact or quality (Thelwall, 2016b, p. 667).

Consequently, citation-based indicators cannot be used at low aggregation levels, as has been widely recognized (Allen, Jones, Dolby, & Walport, 2009; Brito & Rodríguez-Navarro, 2018; Ruiz-Castillo, 2012; Tijssen, Visser, & van Leeuwen, 2002; van-Raan, 2005; Welljams-Dorof-1997 quoted by Aksnes et al., 2019, p. 8). Consistently with this notion, both Pride and Knoth (2018) and Traag and Waltman (2019) have shown that the correlation between peer review and citation-based indicators is higher at an institutional level than for individual publications.

This study investigates the consistency of the *h* index using synthetic series of numbers that simulate citations to the publications from a country or institution. The *h*-index values of these series are compared to the numbers of simulated papers that exceed selected citation thresholds.

## 2. Conceptual meaning of citation thresholds

"While bibliometrics are widely used for research evaluation purposes, a common theoretical framework for conceptually understanding, empirically studying, and effectively teaching its usage is lacking" (Bornmann & Marewski, 2019, p. 419). This situation implies that many research assessment indicators have been proposed without a clear understanding about what they are trying to measure. In other words, most indicators "are largely based on what can easily be counted rather on what really counts" (Abramo & D'Angelo, 2014, p. 1130).

There is nothing against heuristic indicators (Bornmann & Marewski, 2019) as long as they are validated against external criteria (Harnad, 2009). However, until this is done, scientific success cannot be associated with a high value for a bibliometric indicator that has been devised by an imaginative approach.

In contrast with those bibliometric indicators lacking a conceptual understanding, the use of the number of papers that exceed certain citation thresholds is conceptually clear. If, statistically, the number of citations correlates with the scientific impact or quality of a paper (Section 1), it can be concluded that those papers that are more cited are statistically more important than those that receive lower number of citations. Additionally, calculating the proportions of highly cited publications in a field implies a normalization approach that does not apply to the *h* index (Bornmann & Leydesdorff, 2018). However, a clear concept does not always lead directly to the proposal of a reliable indicator; in the case of highly cited publications, the level of citations has to be carefully selected. This is crucial because, in comparing two institutions or countries (A and B), it can be the case that up to a certain level of citations A is ahead of B, but at higher levels of citations B is ahead of A (Brito & Rodríguez-Navarro, 2018). Thus, contradictory comparative performances can be obtained depending on the number of citations at which performance is measured. "Dichotomous procedures rely on the idea that only the upper part of the distribution matters" (Albarrán, Herrero, Ruiz-Castillo, & Villar, 2017, p. 628), but the problem that arises is that, although the number of publications in the two upper parts of two particular citation distributions may be equal, the citations to these two sets of publications can be very different. Consequently, two institutions that are considered equal because they have an equal number of publications attaining a certain citation threshold can be very different if one looks at the distribution of the citations to the two sets of selected publications.

The selection of a certain citation threshold for research assessment—or a top percentile when the number of citations is normalized with reference to global publications (Bornmann & Mutz, 2011; Leydesdorff, Bornmann, Mutz, & Opthof, 2011)—might be considered arbitrary (Schreiber, 2013). This selection, however, should be rational and not arbitrary, depending on the evaluation purposes. The relevant issue is that evaluations can be performed at different scientific levels. For example, in the UK the Research Excellence Framework establishes four quality levels in terms of originality, significance, and rigor: (i) world-leading; (ii) internationally excellent; (iii) recognized internationally; and (iv) recognized nationally (REF2014, 2011, p. 43). Most scientists probably also recognize the Nobel Prize level. Each of these levels corresponds to a certain top percentile of most cited papers. For example, in chemistry the world-leading level corresponds to the top 3% by citation (Rodríguez-Navarro & Brito, 2020), and the Nobel Prize level corresponds with the top 0.01% by citation (Brito & Rodríguez-Navarro, 2018). Thus, the selection of the scientific level, for either a peer-review or a bibliometric indicator in a research evaluation, is a logical requirement in a dichotomous evaluation. Evaluators have to fix this level according to their requirements.

It is worth noting that if research performance is measured at high citation levels, or in terms of the number of publications among global top percentiles, the results will depend on both the total number of publications of the country or institution and the capacity of the research system to produce highly cited papers. At high citation levels, there can be compensation for a lower total number of papers through a higher capacity of the research system to produce highly cited papers. For example, the Massachusetts Institute of Technology (MIT) publishes fewer papers than many countries, but publishes more papers at the Nobel Prize level than many of these countries.

In summary, because the *h* index is a dichotomous size-dependent indicator, if it measures performance it will do so only for a certain citation level; from a formal point of view, it cannot be a universal indicator of research performance. Therefore, to investigate the consistency of this index for research assessment, different levels of citations have to be investigated.

## 3. Aim and methods

This study aims to demonstrate the inconsistencies of the *h* index as an indicator for research assessment and the way in which it leads to misleading results. A previous paper by Waltman and van Eck (2012) has already investigated this matter using examples. Our study addresses the same matter from a mathematical point of view, using the numbers of papers exceeding various citation thresholds as references to which we compare the *h* index.

For these purposes we take advantage of the fact that citations to the publications in a certain research system follow a lognormal distribution (Evans, Hopkins, & Kaube, 2012; Radicchi, Fortunato, & Castellano, 2008; Redner, 2005; Rodríguez-Navarro & Brito, 2018; Stringer, Sales-Pardo, & Amaral, 2010; Thelwall & Wilson, 2014a, 2014b;

Viiu, 2018). Thus, a research system is fully characterized by the number of papers, $N$, and the $\mu$ and $\sigma$ parameters of its lognormal distribution of citations.

In a lognormal distribution the probability density function, denoted by *PDF(c)*, is:

$$PDF(c) = lognormal\ (c; \mu, \sigma) = \frac{1}{\sigma c \sqrt{2\pi}}\ e^{-((lnc-\mu)^2/2\sigma^2)} \quad (1)$$

so the number of papers with $c$ citations, is simply:

$$Number(c) = N \cdot PDF(c) \quad (2)$$

The average number of citations is calculated from Eq. 1 (Aitchison and Brown, 1963),

$$<c> = e^{(\mu+\sigma^2/2)} \quad (3)$$

Thus the total number of citations is simply the number of papers multiplied by the average number of citations:

$$\Sigma(c) = N\ e^{(\mu+\sigma^2/2)} \quad (4)$$

In lognormal distributions, Eq. 1 applies to a continuous distribution. Although the number of citations is a discrete (integer) variable and there are ways to discretize the lognormal distribution, "it is reasonable to use the continuous approximation in order to be able to mathematically analyse citation indicators" (Thelwall, 2016ª, p. 872).

We can calculate from Eq. 1 the probability that a paper gets $c$ or more citations, $P(c)$, and from Eq. 2 the frequency of papers with $c$ or more citations. The expression for this frequency is the cumulative distribution function:

$$F(c) = \int_c^\infty N \cdot PDF(c)dc \quad (5)$$

The integral in (5) can be evaluated to give (Aitchison & Brown, 1963):

$$F(c) = N\left\{\frac{1}{2} + \frac{1}{2}Erf\left[\frac{\mu - \ln(c)}{\sigma\sqrt{2}}\right]\right\} \quad (6)$$

where *Erf* is the error function.

Next, in a continuous approximation, the *h* index is defined as that there are *h* papers that has *h* or more citations. Using that definition, the *h* index can be calculated from *F(c)* as:

$$F(h) = h \quad (7)$$

The solution of Eq. 7, which defines *h* as a function of *N*, $\mu$, and $\sigma$, cannot be obtained analytically; it has to be calculated by numerical methods. However, it is possible to obtain an asymptotic expansion when *N*, and therefore *h*, are large, by expanding the *Erf(h)* function that appears in Eq. 6 (Gautschi, 1965). The result is:

$$\log h \approx \sigma \sqrt{2 \log N} \quad (8)$$

This relation shows the nonlinear relationship between *N* and *h*. Moreover, it implies a slow growth of *h* with the number of publications *N*.

Together with the *h* index, we also computed the number of papers with numbers of citations equal to or surpassing 5, 10, 20, 50, 100, and 500 citations. It is simple to do this from the analytical description of *PDF(c)* in terms of the lognormal distribution; for example for 5 citations:

$$F(5) = \int_5^\infty N \cdot PDF(c)\, dc \quad (9)$$

Similar expressions apply for the other citation thresholds.

An alternative approach to the analytical method based on Eq. 1–9 consists in the generation of series of numbers simulating the numbers of citations received by the papers published by a country or institution. To create these series, we generated random numbers that are lognormal distributed; in order to avoid the intrinsic fluctuations inherent in the generation of random numbers, we generated 10,000 lognormal distributions for each parameter set, and averaged them to obtain the final data points. Next, the continuous lognormally distributed numbers were discretized to create a discrete lognormal distribution of citations (Rodríguez-Navarro & Brito, 2018; Thelwall, 2016c), and we ranked these from highest to lowest. For simplicity, in the rest of this study, we will refer to these simulated series of papers and citations as papers and citations, omitting the fact that they are synthetic series.

To study the $h$ index in a wide range of conditions, the $\mu$ and $\sigma$ parameters were fixed at values within the extreme values found in real evaluations for a single research field and a defined citation window. Both circumstances affect the $\mu$ and $\sigma$ parameters because they affect the number of citations. $N$ varied from 200 to 10,000; this 50-fold variation probably is not exceeded in country or institution comparisons. $\mu$ varied from 2.7 to 1.3 and the values of $\sigma$ were varied depending on the values of $\mu$ (Rodríguez-Navarro & Brito, 2018). The highest values, $\mu = 2.7$ and $\sigma = 1.2$, correspond to the distribution of citations in a field such as chemistry in a highly competitive institution such as MIT in a citation window of four years. In the same conditions a top European research university has $\mu = 2.2$ and $\sigma = 1.1$ (unpublished results, but see Rodríguez-Navarro & Brito, 2020). The lowest values, $\mu = 1.3$ and $\sigma = 0.8$, correspond to a below-average country or institution. We selected 30 combinations of parameters, which should reasonably be sufficient to demonstrate the consistency or inconsistency of the $h$ index.

Table 1 shows the three defining parameters of the 30 lognormal citation series: $\mu$, $\sigma$, and number of papers. It also shows the sum of citations, the values of the $h$ index, and the probabilities of papers with numbers of citations that are equal to or exceed the already mentioned six thresholds for citations ($P(x)$; $x = 5, 10, 20, 50, 100,$ and $500$).

< Table 1 about here >

In the synthetic series, the *h* index and the total number of citations can be manually computed, independently of whether *N* is large or small. In contrast, the number of papers with citations equal to or above a certain number of citations can be manually computed depending on *N*; if *N* is small, the number of papers with a high number of citations might be zero. Therefore, in many cases we used a calculation method from the lognormal parameters (Section 3) and computed fractional numbers. Using this method we avoid using very large series to investigate differences between series with very different $\mu$ and $\sigma$ parameters. When manual computation is possible, the results from this method and the analytical method are equal.

## 4. Results

### 4.1. Dependence of the *h* index on the number of papers

The *h*-index value given by Eq. 6 and 7 predicts a complex dependence of the *h* index on the total number of papers, *N*. Figure 1 shows three examples of the variation of *h*, for series with $\mu$ and $\sigma$ parameters that are the highest, middle, and lowest values in Table 1, calculated by numerical analysis. As expected, the *h* index does not grow proportionally with the number of papers. Except when there is a very low number of papers (e.g. $N < 30$), *h* grows visibly faster than number of papers; as the number of papers increases the growth rate of *h* decreases and at a certain number of papers the rate becomes almost zero. It is worth noting that Eq. 8 obtained by asymptotic approximation reproduces quite accurately the analytical curves (thinner lines in Figure 1).

< Figure 1 about here >

The higher the $\mu$ of the citation distribution, the wider the range of the number of papers within which *h* increases visibly. In other words, an increase in the number of papers increases the *h* index more in the most efficient research systems, with the highest $\mu$, than in less competitive systems. For example, in the three examples presented in Figure 1, the changes in the *h*-index values when the number of papers increases from 100 to 200 are the following: $\mu = 2.7$, $\sigma = 1.2$, from 29 to 41; $\mu = 2.1$, $\sigma = 1.1$, from 20 to 28; $\mu = 1.3$, $\sigma = 0.8$, from 10 to 13.

**4.2. *h* index versus the number of papers at different citations thresholds**

Because the *h* index is a size-dependent indicator, the second step in our study was to investigate the correlation of the *h* index with the number of papers that exceeds the six citation thresholds selected in our study—these numbers were obtained by multiplying *N* by the corresponding probability, *P*(*x*), in Table 1. Figure 2 shows the scatter plots of the *h*-index values as functions of the number of papers exceeding the six citation thresholds. Visual inspection of these plots reveals that the *h*-index value shows a nonlinear correlation with the number of successful papers. The correlation is poor at low, *F*(5) and *F*(10), and high, *F*(500), citation levels, but better at intermediate, *F*(20), *F*(50), and *F*(100), citation levels. For these citation thresholds, especially for *F*(100), the data can be reasonably fitted by a power law.

< Figure 2 about here >

If the relationship between the *h* index and *F*(50) or *F*(100) were a perfect power law, the *h* index could be transformed into a corrected indicator exactly proportional to *F*(50) or *F*(100), which is an added value for an indicator, by applying a simple power law transformation. However, although the goodness of fit to power laws of the *F*(50) and *F*(100) values versus the *h*-index values for the 30 series are high ($R^2 = 0.98$ and 0.97, respectively), when they are examined at low *h*-index values the deviations are high (Figure 3). These deviations imply that the *h* index is not a good indicator for *F*(50) or *F*(100) at low *h*-index values ($h < 30$).

< Figure 3 about here >

**4.3. *h/N* versus probabilities at different citation levels**

A reliable size-dependent indicator should be transformed into a size-independent indicator by dividing it by the size of the system. In the case of the *h* index, its complex dependence on *N* suggests that any relationship between *h* and *F*(*x*) would be lost by dividing by *N*. Figure 4 shows the six scatter plots of *h*/*N* versus the probabilities recorded in Table 1; visual inspection of these plots roundly demonstrates that *h*/*N* does not correlate with the probability of a research system publishing papers that exceed a certain citation threshold.

< Figure 4 about here >

**4.4. Is the *h* index a better indicator than the number of citations?**

Many studies, including the original paper proposing the *h* index (Hirsch, 2005), have claimed that the *h* index can be calculated from the total number of citations (see (Malesios, 2015). Specifically, van Raan (2006) found that a power law links the *h* index with the sum of the number of citations. Consistent with this observation, we also observe good fits of our data for Σ(*c*) and *h* to a power law (not shown); the exponent 0.42 is not very different from the exponent given by van Raan (2006), which is 0.45.

These findings raise the question of whether, even if the *h* index was a good research indicator, it would be a better indicator than standard bibliometric measures (Bornmann, Mutz, & Daniel, 2009) and, specifically, whether it would be better than the total number of citations, Σ(*c*).

To answer this question from a size-dependent perspective, we plotted Σ(*c*) as a function of the number of papers exceeding the six citation thresholds in the 30 synthetic series; Figure 5 shows these scatter plots. A visual inspection of the scatter plots shows that the number of citations is linearly correlated with the number of papers exceeding three citation thresholds, for 10, 20, and 50 citations. The lowest data point deviations occur at *F*(10).

< Figure 5 about here >

As in the case of the *h* index the next step was to study the correlation between the size-independent indicator Σ(*c*)/*N* and the probability of receiving at least a certain number of citations. Figure 6 shows the scatter plots for the six citation thresholds; visual inspection of the scatter plots reveals that a transformation into a size-independent indicator improves the correlations, namely the transformation into a size independent indicator greatly decreases the data point deviations from the general trend. The scatter plots change from concave at low citation thresholds to convex at high citation thresholds. For a threshold of 20 citations, the linear correlation, although visibly not perfect, is very high (the Pearson correlation coefficient between *P*(20) and Σ(*c*)/N is 0.988 with a *p* value of $10^{-24}$), and it increases slightly for *P*(30) (Pearson correlation coefficient of 0.998). Thus we can write:

$$\Sigma(c)/N \approx 4.9 + 88.7\, P(30) \qquad (10)$$

In summary, in contrast to *h*/*N* (Fig. 4), the correlation between the mean number of citations (Σ(*c*)/N) and the probability of publishing a paper exceeding a fixed threshold of between 10 and 100 citations is very high. The correlation is almost linear for the 20 and 30 citation thresholds.

< Figure 6 about here >

## 5. Discussion

Citations of papers from institutions or countries are lognormally distributed, and we performed our study using 30 series of numbers that were lognormally distributed with *μ* and *σ* parameters that were equal to the real values found in the citation distributions of countries and institutions (Section 3). The first observation that questions the validity of the *h* index is its atypical dependence on the number of publications (Figure 1). This dependence has been extensively studied (Egghe & Rousseau, 2006; Glänzel, 2006; Malesios, 2015; Molinari & Molinary, 2008; Montazerian, Zanotto, & Eckert, 2019);

previous and our present results have many points in common, but a comparative study is out of the scope of this report.

It can be taken for granted that any institution or country at any level of success that doubles its number of papers should also double its research performance. Otherwise, it would have to be the case that when the number of papers increases, the probability of the success of each paper decreases; this cannot be accepted as a property for all research systems. In other words, in institutions that are functionally identical (e.g. Series 12–15), the size-dependent indicators have to be proportional to the number of papers. The *h* index does not fulfill this simple principle and only for this reason it should be rejected as a reliable research indicator, and there are also other reasons for its rejection.

The *h* index is a size-dependent indicator that is too simple to reveal all the characteristics of a research system, but it may reveal the number of papers that surpass a certain level of scientific relevance. Therefore, we took as benchmarks the number of papers that surpass certain numbers of citations. Although individually a paper can receive a greater or smaller number of citations than it deserves according to the importance of the science it reports, when many papers are aggregated, the number of citations of these papers should reveal their joint importance (Section 1). The rationale of the approach is that the target of research in all institutions and countries is to make research that propels the advancement of science at the highest possible level, which means that highly cited papers should be the target of research (Section 2). Basically, it can be said that, so far, "the benefits of scientific discoveries have been heavy-tailed" (Press 2013, p. 1130). This rationale, however, does not fix the citation level that should be exceeded.

Our results clearly demonstrate that the *h* index is not an indicator of the number of papers that exceeds a certain threshold of citations from as few as five to as many as 500 citations. For the 50 and 100 citation thresholds, there is an apparent power law correlation between the *h* index and the number of papers (Figure 2). Such nonlinear correlation is not convenient because the comparisons of institutions or countries are not the same at high and low *h*-index values. This problem alone could be easily overcome by transforming the *h* index into a more convenient linearly correlated index by a simple power law transformation, but the numerous data point deviations from the general trend cannot be overcome (Figure 3).

As a consequence of its atypical dependence on the number of papers, if we divide by the number of papers, *h/N*, and compare the results with probabilities, any vestige of correlation disappears (Figure 4). For example, the *h*-index values for series 13, 20, and 30 are very similar 27, 27, and 28, for which the total numbers of publications are 200, 500, and 5000, respectively (Table 1). When we divide by the number of papers, the data points of the three series that are very close in Figure 2 become very distant in Figure 4. Because this dispersion applies to all data points, the apparent correlation between *h* and *F*(100) disappears when the two values are divided by *N*—there is no correlation between *h/N* and *P*(100).

Furthermore, the lack of consistency of the *h* index as a research performance indicator arises more clearly when it is compared with the total number of citations, $\Sigma(c)$. A simple visual inspection of the scatter plots (Figures 2 and 5) demonstrates the superiority of $\Sigma(c)$. In fact, $\Sigma(c)$ is not a bad indicator of *F*(10). More interestingly, the transformation of $\Sigma(c)$ into a size-independent indicator, $\Sigma(c)/N$, improves the correlation (Figure 5). Although the correlation is never linear, for 20 and 30 citations the sigmoid correlations approximate very closely to linear correlations. It can be concluded that the mean number of citations is a good indicator of the probability that a publication of an institution or country will exceed the 20 or 30 threshold of citations. Thus, from a mathematical point of view, Eq. 10 is a basic equation in scientometrics. In real cases, however, there might be deviations, because occasionally several publications are big hits that deviate from the lognormal distribution and increase the mean number of citations.

One proposed advantage of the *h* index over the total number of citations is that the former is not "inflated by a small number of big hits" (Hirsch, 2005), p. 16569). This advantage would be real if the *h* index measured the success or performance of research in any way, but this has never been demonstrated (Section 1), and our study demonstrates just the opposite. Assuming that citations to the publications from countries and institutions follow lognormal distributions, as stated by Bornmann (2014), the use of *h*-index in research evaluations, like that of the Journal Impact Factor, should disappear.

Acknowledgments

This work was supported by the Spanish Ministerio de Economía y Competitividad (Grant number FIS2017-83709-R).

Table 1. Description of the 30 lognormal functions used in this study, including the total number of citations (Σ(*c*)), *h* index, and probabilities that the number of citations of a random paper in the series is equal or surpassing 5, 10, 20, 50, 100, and 500 citations

| Series | μ | σ | N | Σ(c) | h | P(5) | P(10) | P(20) | P(50) | P(100) | P(500) |
|---|---|---|---|---|---|---|---|---|---|---|---|
| 1 | 2.7 | 1.2 | 500 | 15280 | 61 | 0.8183 | 0.6297 | 0.4027 | 0.1562 | 0.0562 | 1.70E-03 |
| 2 | 2.7 | 1.2 | 5000 | 152891 | 145 | 0.8183 | 0.6297 | 0.4027 | 0.1562 | 0.0562 | 1.70E-03 |
| 3 | 2.7 | 1.2 | 1000 | 30593 | 80 | 0.8183 | 0.6297 | 0.4027 | 0.1562 | 0.0562 | 1.70E-03 |
| 4 | 2.5 | 1.2 | 500 | 12510 | 54 | 0.7710 | 0.5653 | 0.3398 | 0.1197 | 0.0397 | 9.82E-04 |
| 5 | 2.5 | 1.1 | 4000 | 89223 | 104 | 0.7909 | 0.5712 | 0.3261 | 0.0996 | 0.0278 | 3.67E-04 |
| 6 | 2.4 | 1.1 | 500 | 10093 | 47 | 0.7638 | 0.5353 | 0.2941 | 0.0846 | 0.0225 | 2.62E-04 |
| 7 | 2.3 | 1.1 | 5000 | 91364 | 97 | 0.7349 | 0.4991 | 0.2635 | 0.0714 | 0.0181 | 1.86E-04 |
| 8 | 2.3 | 1.1 | 1000 | 18266 | 57 | 0.7349 | 0.4991 | 0.2635 | 0.0714 | 0.0181 | 1.86E-04 |
| 9 | 2.3 | 1.1 | 10000 | 182662 | 120 | 0.7349 | 0.4991 | 0.2635 | 0.0714 | 0.0181 | 1.86E-04 |
| 10 | 2.2 | 1.1 | 3000 | 49564 | 77 | 0.7043 | 0.4628 | 0.2347 | 0.0598 | 0.0144 | 1.31E-04 |
| 11 | 2.2 | 1.1 | 10000 | 165273 | 112 | 0.7043 | 0.4628 | 0.2347 | 0.0598 | 0.0144 | 1.31E-04 |
| 12 | 2.1 | 1.1 | 500 | 7483 | 39 | 0.6722 | 0.4269 | 0.2077 | 0.0497 | 0.0114 | 9.18E-05 |
| 13 | 2.1 | 1.1 | 200 | 2987 | 27 | 0.6722 | 0.4269 | 0.2077 | 0.0497 | 0.0114 | 9.18E-05 |
| 14 | 2.1 | 1.1 | 5000 | 74771 | 84 | 0.6722 | 0.4269 | 0.2077 | 0.0497 | 0.0114 | 9.18E-05 |
| 15 | 2.1 | 1.1 | 10000 | 149541 | 104 | 0.6722 | 0.4269 | 0.2077 | 0.0497 | 0.0114 | 9.18E-05 |
| 16 | 2.1 | 1.0 | 3000 | 40390 | 63 | 0.6881 | 0.4197 | 0.1852 | 0.0350 | 0.0061 | 1.94E-05 |
| 17 | 2.0 | 1.0 | 3000 | 36545 | 58 | 0.6519 | 0.3811 | 0.1597 | 0.0279 | 0.0046 | 1.25E-05 |
| 18 | 1.9 | 1.0 | 5000 | 55116 | 63 | 0.6143 | 0.3436 | 0.1366 | 0.0221 | 0.0034 | 7.99E-06 |
| 19 | 1.9 | 1.0 | 1000 | 11026 | 39 | 0.6143 | 0.3436 | 0.1366 | 0.0221 | 0.0034 | 7.99E-06 |
| 20 | 1.7 | 1.0 | 500 | 4520 | 27 | 0.5361 | 0.2734 | 0.0975 | 0.0135 | 1.84E-03 | 3.17E-06 |
| 21 | 1.7 | 1.0 | 300 | 2709 | 23 | 0.5361 | 0.2734 | 0.0975 | 0.0135 | 1.84E-03 | 3.17E-06 |
| 22 | 1.7 | 1.0 | 100 | 901 | 15 | 0.5361 | 0.2734 | 0.0975 | 0.0135 | 1.84E-03 | 3.17E-06 |
| 23 | 1.7 | 0.9 | 2000 | 16411 | 36 | 0.5401 | 0.2516 | 0.0750 | 0.0070 | 6.23E-04 | 2.63E-07 |
| 24 | 1.5 | 0.9 | 500 | 3360 | 21 | 0.4516 | 0.1863 | 0.0483 | 0.0037 | 2.80E-04 | 8.10E-08 |
| 25 | 1.5 | 0.9 | 200 | 1344 | 16 | 0.4516 | 0.1863 | 0.0483 | 0.0037 | 2.80E-04 | 8.10E-08 |
| 26 | 1.5 | 0.9 | 2000 | 13437 | 31 | 0.4516 | 0.1863 | 0.0483 | 0.0037 | 2.80E-04 | 8.10E-08 |
| 27 | 1.5 | 0.9 | 3000 | 20161 | 35 | 0.4516 | 0.1863 | 0.0483 | 0.0037 | 2.80E-04 | 8.10E-08 |
| 28 | 1.4 | 0.9 | 1000 | 6084 | 24 | 0.4080 | 0.1580 | 0.0381 | 0.0026 | 1.85E-04 | 4.41E-08 |
| 29 | 1.3 | 0.8 | 1000 | 5060 | 19 | 0.3495 | 0.1051 | 0.0170 | 5.47E-04 | 1.80E-05 | 4.04E-10 |
| 30 | 1.3 | 0.8 | 5000 | 25272 | 28 | 0.3495 | 0.1051 | 0.0170 | 5.47E-04 | 1.80E-05 | 4.04E-10 |

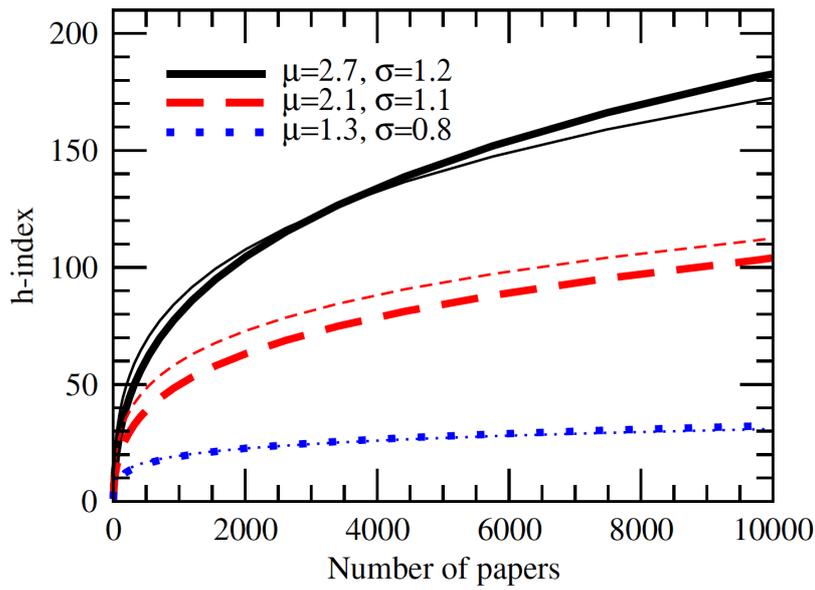

Figure 1. Response of the h index to the increase of the number of publications when the number of citations obeys a lognormal distribution. The thicker line corresponds to the solution of the equation that defines h. The thinner line corresponds to the simplified Eq. 8.

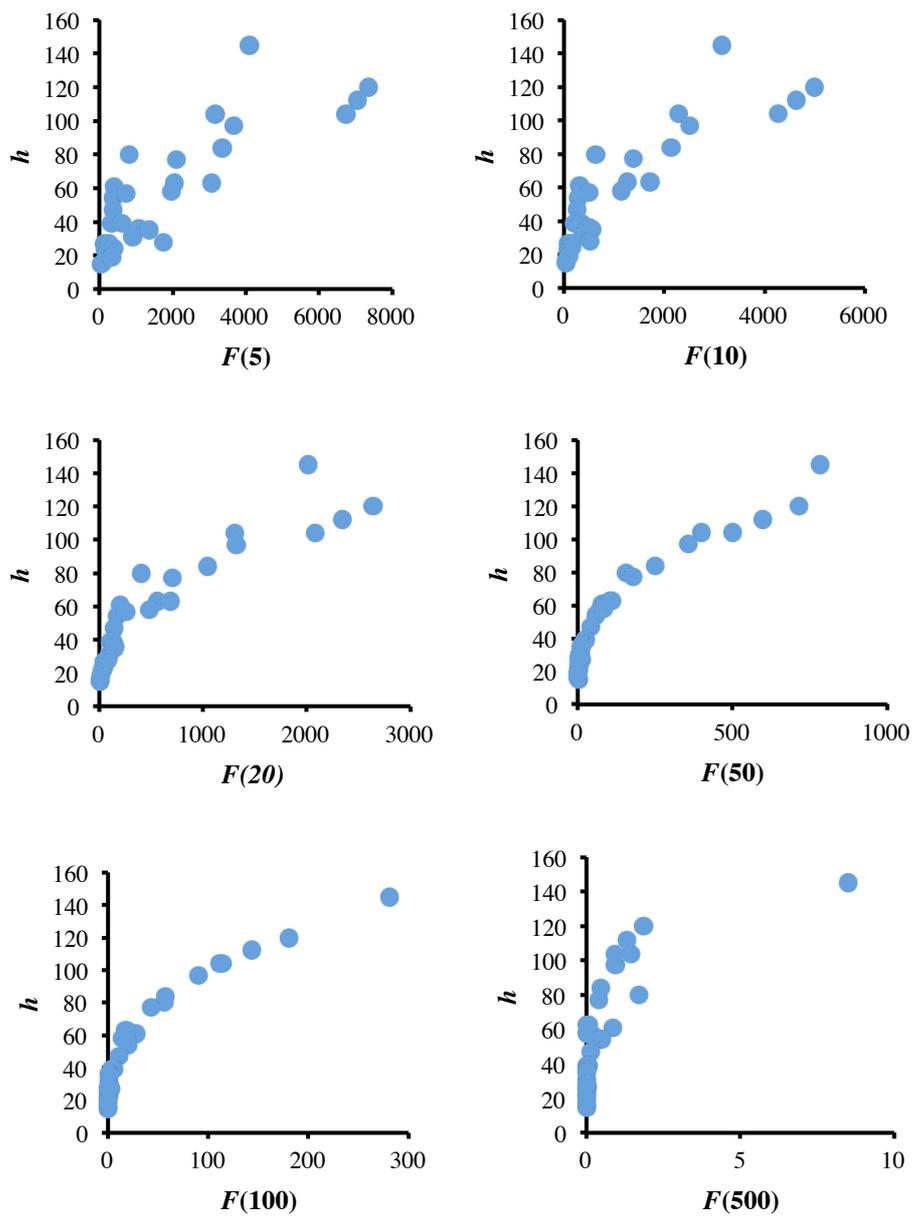

Figure 2. Scatter plot of the h index versus the number of publications that exceed six citation thresholds, from 5 to 500 citations. Data points of the 30 simulated citation series recorded in Table 1

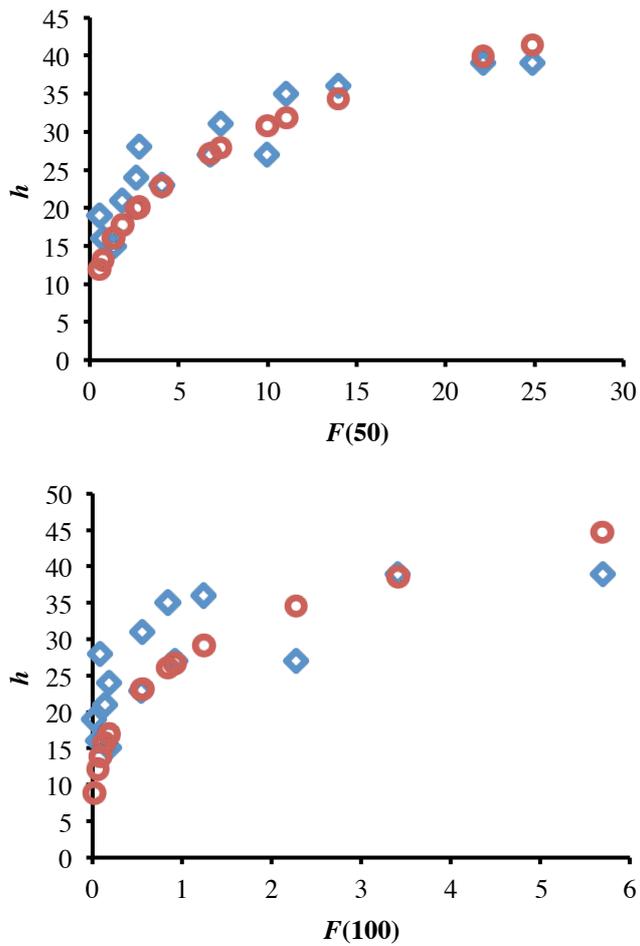

Figure 3. Scatter plot of the h index versus the number of publications that exceed the 50 and 100 citation thresholds. Symbols: Diamonds, real-value data points, as in Figure 2; circles, computed values that were obtained from the power law equations that best fit the 30 data points in Figure 2: $h = 14.6\, F(50)^{0.325}$ and $h = 27.4\, F(100)^{0.282}$

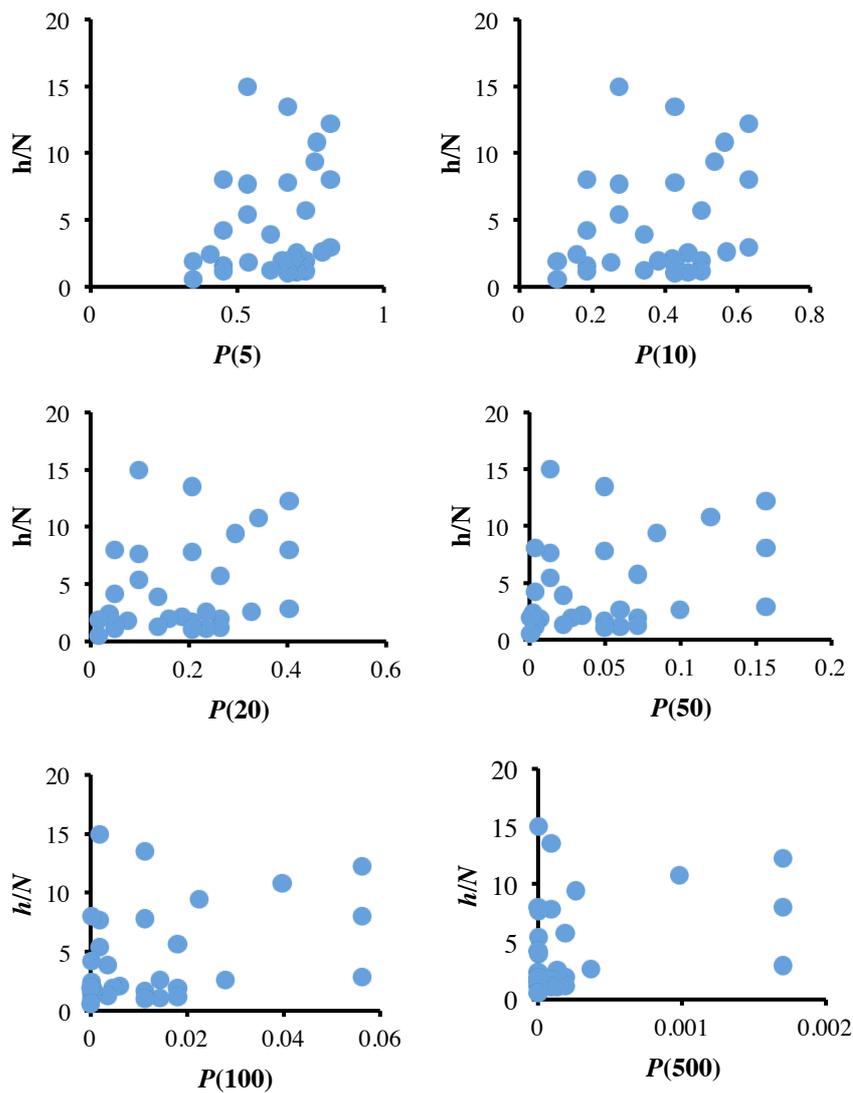

Figure 4. Scatter plot of h/N versus the probability that a random paper exceeds six citation thresholds, from 5 to 500 citations. Data points as in Figure 2 but dividing by N the h-index and frequency values

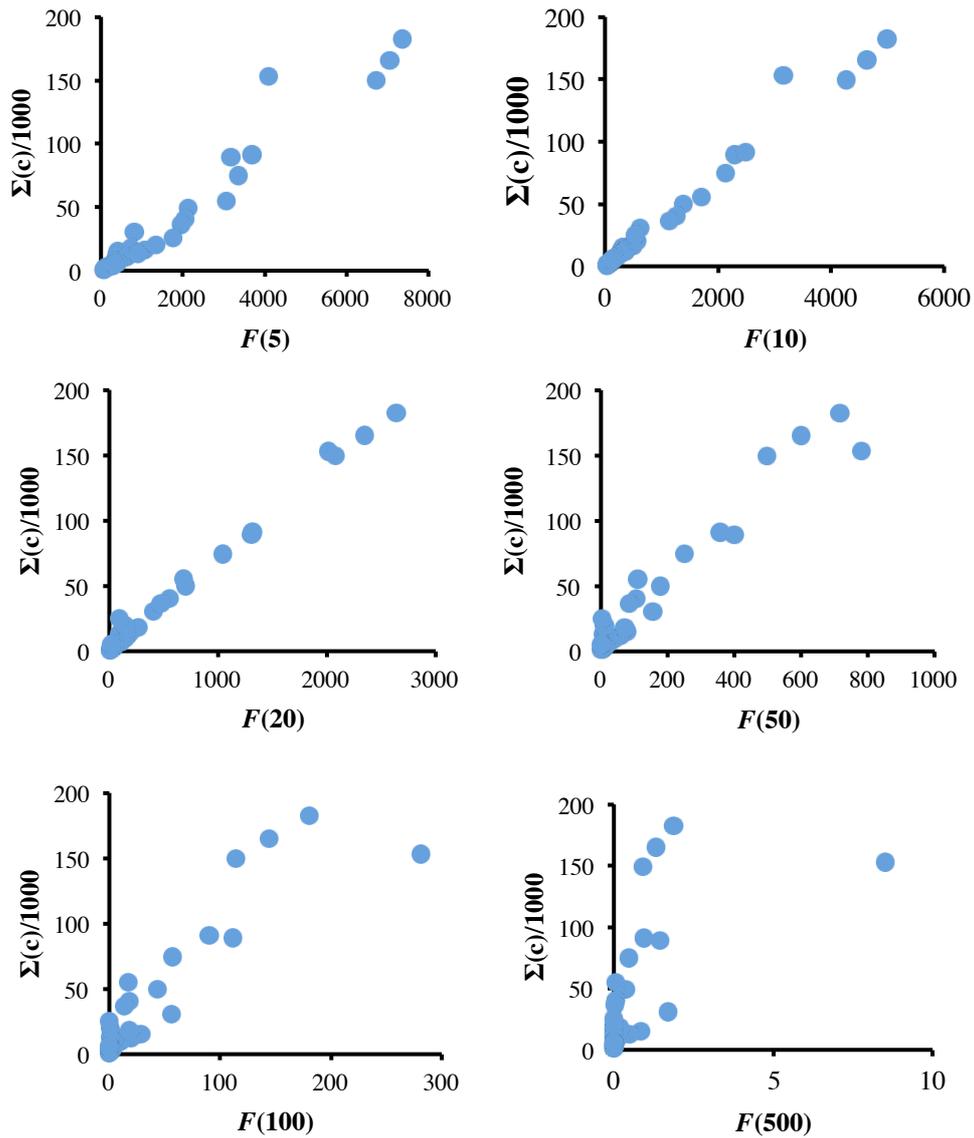

Figure 5. Scatter plots of the total number of citations ($\Sigma(c) \cdot 10^{-3}$) versus the number of publications that exceed six citation thresholds, from 5 to 500 citations. Data points of the 30 simulated citation series recorded in Table 1

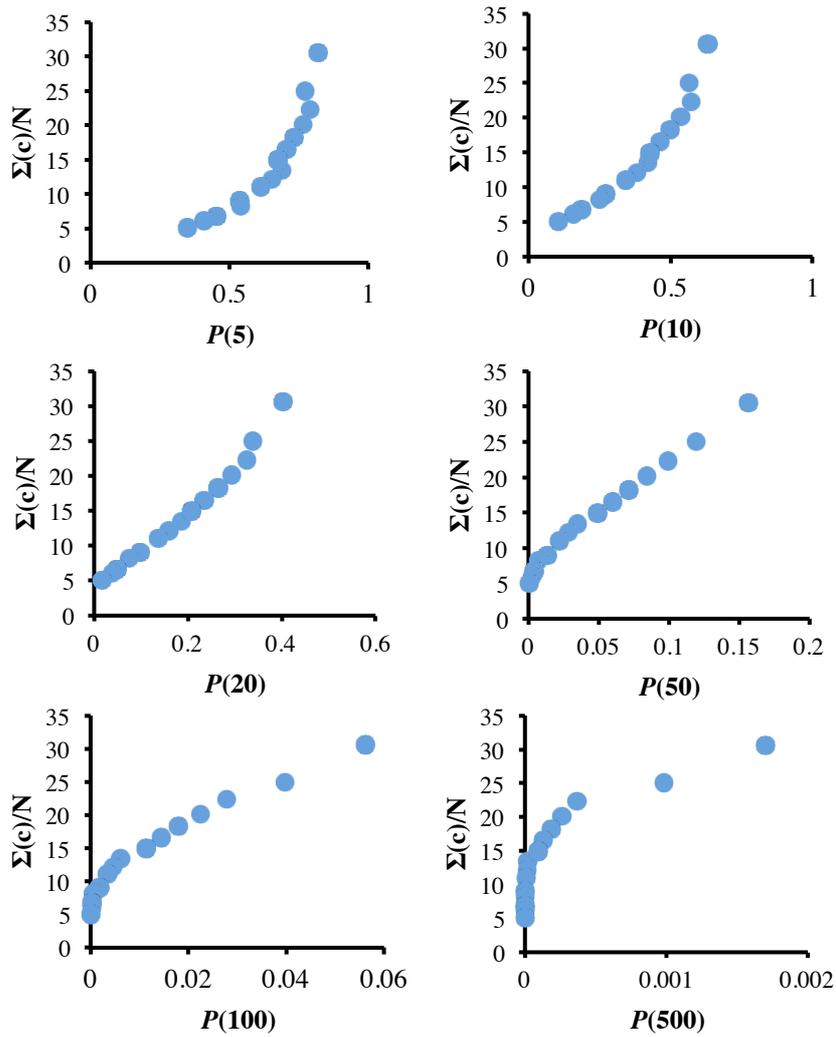

Figure 6. Scatter plot of Σc/N versus the probability that a random paper exceeds six citation thresholds, from 5 to 500 citations. Data points as in Figure 5 but dividing by N the Σc and frequency values